\documentclass{emulateapj}

\newcommand{\kms}{\ensuremath{\mathrm{km~s}^{-1}}}

\newcommand{\msun}{\ensuremath{M_\odot}}

\newcommand{\Code}{\texttt{SEDONA}}
\newcommand{\Nifs}{\ensuremath{^{56}\mathrm{Ni}}}
\newcommand{\Cofs}{\ensuremath{^{56}\mathrm{Co}}}

\newcommand{\texp}{\ensuremath{t_{\mathrm{exp}}}}
\newcommand{\dmf}{\ensuremath{\Delta m_{15}}}
\newcommand{\dmfb}{\ensuremath{\Delta m_{15}(B)}}
\newcommand{\nuc}[2]{\ensuremath{\mathrm {^{#2}#1}}}

\shortauthors{Kasen \& Plewa}
\shorttitle{Observational Properties of DFD}

\begin{document}

\title{Detonating Failed Deflagration Model of Thermonuclear Supernovae II. Comparison to Observations}

\author{Daniel Kasen\altaffilmark{1,2}, Tomasz Plewa\altaffilmark{3,4}}
\altaffiltext{1}{Allan C. Davis Fellow, Department of Physics and Astronomy, Johns Hopkins University,
Baltimore, MD 21218}
\altaffiltext{2}{Space Telescope Science Institute, Baltimore, MD 21218}
\altaffiltext{3}{Center for Astrophysical Thermonuclear Flashes,
   University of Chicago,
   5640 South Ellis Avenue,
   Chicago, IL 60637}
\altaffiltext{4}{Department of Astronomy \& Astrophysics,
   University of Chicago,
   5640 South Ellis Avenue,
   Chicago, IL 60637}
\begin{abstract} 
We develop and demonstrate the methodology of testing
multi-dimensional supernova models against observations by studying
the properties of one example of the detonation from failed
deflagration (DFD) explosion model of thermonuclear supernovae.  Using
time-dependent multi-dimensional radiative transfer calculations, we
generate the synthetic broadband optical light curves, near-infrared
light curves, color evolution curves, full spectral time-series, and
spectropolarization of the model, as seen from various viewing angles.
All model observables are critically evaluated against examples of
well-observed, standard Type~Ia supernovae (SNe~Ia).  We explore the
consequences of the intrinsic model asphericity by studying the
dependence of the model emission on viewing angle, and by quantifying
the resulting dispersion in (and internal correlations between)
various model observables.  These
\emph{statistical} properties of the model are also evaluated against those
of the available observational sample of SNe~Ia.  On the whole, the
DFD model shows good agreement with a broad range of SN~Ia
observations.  Certain deficiencies are also apparent, and point to
further developments within the basic theoretical framework.  We also
identify several intriguing orientation effects in the model which
suggest ways in which the asphericity of SNe~Ia may contribute to
their photometric and spectroscopic diversity and, conversely, how the
relative homogeneity of SNe~Ia constrains the degree of asymmetry
allowable in the models.  The comprehensive methodology adopted in
this work proves an essential component of developing and validating
theoretical supernova models, and helps motivate and clearly define
future directions in both the modeling and the observation of SNe~Ia.
\end{abstract}
\keywords{hydrodynamics --- radiative transfer --- spectropolarimetry
--- supernovae:general}
\section{Introduction}\label{s:introduction}

The critical assessment of theories of thermonuclear, or Type~Ia,
supernova (SN) explosions must ultimately involve not only the
modeling of the complex 3-dimensional physics characterizing the
explosion process itself, but also a comprehensive treatment of the
multi-dimensional radiative transfer physics determining the emission
from the ejected stellar material over the months and years following
explosion, thereby allowing the model to be tested by direct
comparison to a broad range of supernova observations, such as light
curves, spectra and spectropolarization, the aim being to reproduce
the essential features of the observational sample, within the
framework of a certain theoretical paradigm, and with the minimum
number of free parameters and assumptions.

Most theories of Type~Ia supernova (SN~Ia) explosions begin with the
assumption that these events represent the thermonuclear disruption of
carbon-oxygen white dwarfs, typically near the Chandrasekhar limit.
Despite several uncertainties concerning the structure of the
progenitor star and the precise conditions leading to its ignition,
complex multi-physics codes have been used to simulate the
hydrodynamics and turbulent nuclear combustion leading to the
unbinding of the white dwarf.  Nuclear burning typically lasts for up
to 5~sec after ignition; the stellar material ejected in the explosion
reaches the free-expansion phase $\sim 1$ minute later.  At this time,
the model output (i.e., the density, velocity, and compositional
structure of the SN ejecta) may be fed into detailed radiative
transfer codes for calculation of the synthetic model spectra, light
curves and polarization.

The diagnostic value of such transfer calculations in assessing the
model predictions is impressive.  Light curve observations constrain
the explosion energy, the total ejected mass, and the amount and
distribution of the synthesized radioactive products (in particular
\Nifs) which power the SN light curve.  Spectroscopic observations
constrain the ejecta velocity scale, thermal state and chemical
stratification.  Spectropolarization observations constrain
asymmetries and inhomogeneities in the ejecta density and
compositional structure.  Given the extreme complexity of the
underlying phenomenon, no single measure can be used to establish the
viability of a SN explosion model, rather one must consider this broad
set of model observables as a whole.

Inevitably, the comparison of model and observation involves a number
of important uncertainties, including the physical approximations and
geometrical simplifications adopted in the theoretical calculations,
as well as errors in the observations of SNe, their distances and the
degree of dust extinction.  Nevertheless, radiative transfer studies
of 1-dimensional (1D) SN~Ia models have already provided important
insights into the nature and theory of these objects.  
\cite{Hoeflich_Khokhlov_LC}, for example, synthesized light curves for a wide range
of 1D theoretical models, including deflagrations, pulsed detonations
and delayed-detonations, and compared the results directly to
observations.  The spectroscopic properties of the parameterized 1D
deflagration model w7 of \cite{Nomoto_w7} have been frequently
evaluated against observations
\citep{Branch_w7,Harkness_w7,Jeffery_91T, Nugent_hydro, Lentz_94D,
Baron_94D}.  In addition, the synthetic light curves and spectra of 1D
delayed-detonation models have been studied in great detail, both in
the optical and near infrared \citep{Hoeflich_94D, Hoeflich_DD,
Hoeflich_99by, Wheeler_IR}.  All of these studies, by specifying the
particular successes and failures of the models, have helped define
the basic features required of any viable theory of SNe~Ia.  To date,
however, there does not exist a self-consistent, fully unparameterized
SN~Ia model that satisfies all the observable properties of the class.

With the advance of computing power, more realistic explosion models
of SNe~Ia have shown that multi-dimensional effects likely play an
essential role in the explosion process
\citep[e.g.,][]{Khokhlov_1994,Reinecke_3D}. Corresponding advances in the observations, in
particular polarization measurements, have given direct evidence that
the ejecta of SNe~Ia are not entirely spherically symmetric
\citep{Howell_99by,Wang_01el,Leonard_SNIa,Chornock_05hk}.  
The validation of SN models faces new and interesting challenges once
multi-dimensional scenarios are considered.  Although one can always
study the gross, angle-averaged properties of aspherical models using
1D transfer codes
\citep[e.g.,][]{Blinnikov_3D}, ultimately multi-dimensional
transfer calculations are needed to retain the full geometry of the
problem.  In that way we can study how the model properties depend on
viewing angle, and quantify the resulting dispersion in (and internal
correlations between) the various model observables.  Such
\emph{statistical} properties of the models may then also be 
evaluated against those of the available observational sample of
SNe~Ia.

In this paper, we develop and demonstrate such a methodology for
testing multi-dimensional supernova models by studying the observable
properties of one example of the (axisymmetric)
detonating-failed-deflagration (DFD) model of SNe~Ia \cite[][hereafter
Paper~I]{Plewa_DFD}.  Using the multi-dimensional time-dependent
radiative transfer code \Code, we calculate the full time evolution
(day 2 to day 80 after explosion) of the emergent model spectrum as
seen from various viewing angles.  Broadband $UBVRIJHK$ light curves
are then generated by convolving the spectral series with the
appropriate filter transmission profiles.  The intrinsic asymmetry of
the model is studied by calculating synthetic spectropolarimetry and
by examining the dependence of all model observables on
orientation. The high spectral resolution of the calculations allows
us to study individual absorption features in the synthetic spectra,
and in particular the level of line polarization and the temporal
evolution of line Doppler shifts.  All model properties are critically
contrasted with representative examples of normal SNe~Ia.

Our approach constitutes a comprehensive, end-to-end simulation of the
supernova evolution -- from the first moments of ignition, through the
weeks and months following the explosion -- without resorting to any
ad-hoc manipulation of model parameters.  This methodology provides
the most thorough means of evaluating the success and failures of the
particular explosion paradigm, and offers important guidance for
further development of the theory.

\section{DFD Y12 Explosion Model}
\label{s:y12}

\begin{figure*}[ht]
\begin{center}
\includegraphics[height=7.0cm,clip=true]{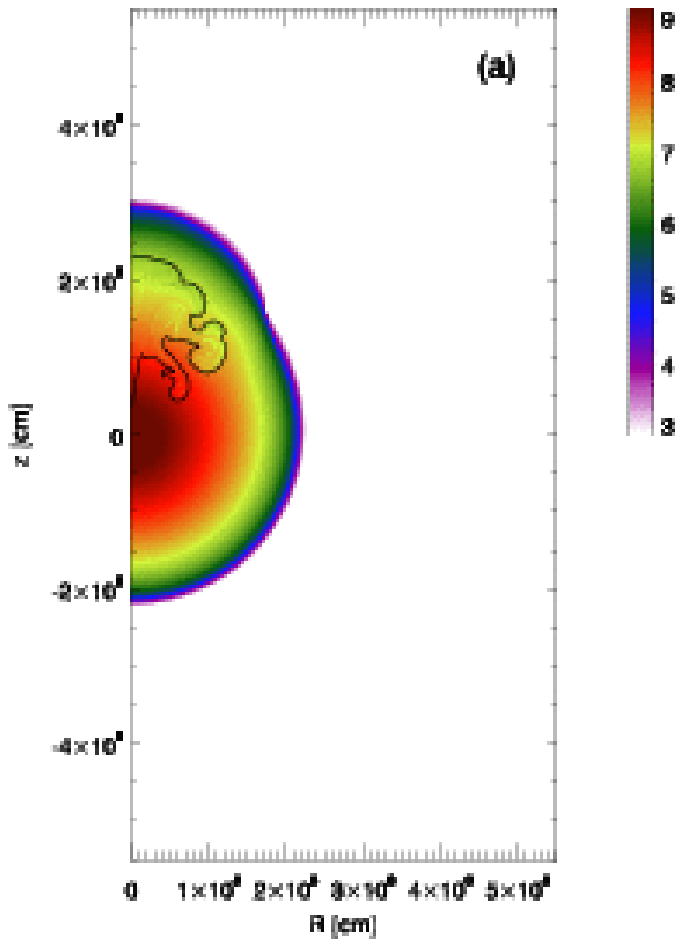}
\includegraphics[height=7.0cm,clip=true]{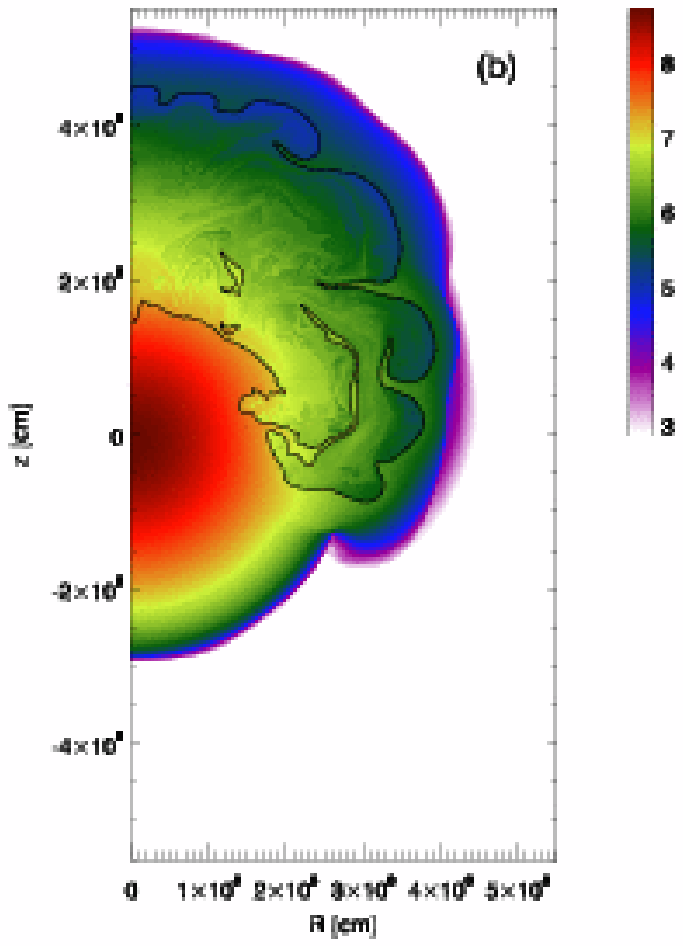}
\includegraphics[height=7.0cm,clip=true]{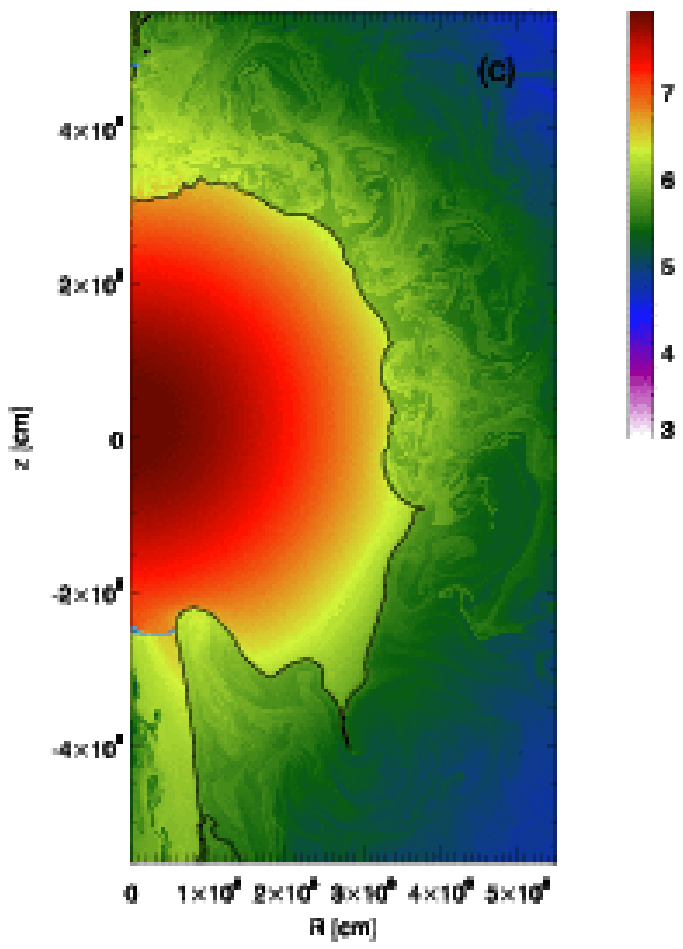}
\includegraphics[height=7.0cm,clip=true]{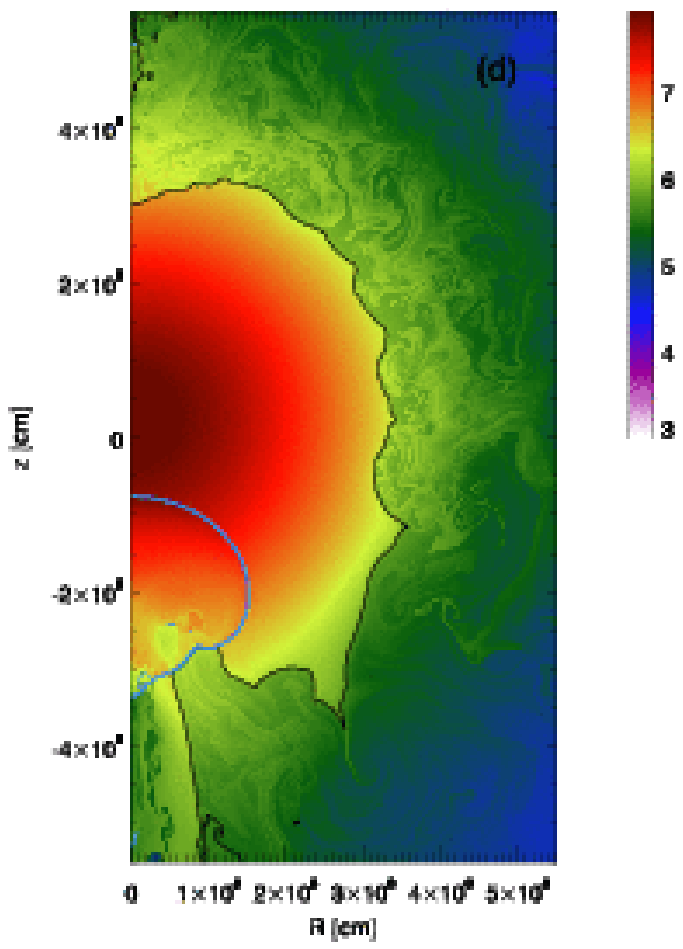}
\includegraphics[height=7.0cm,clip=true]{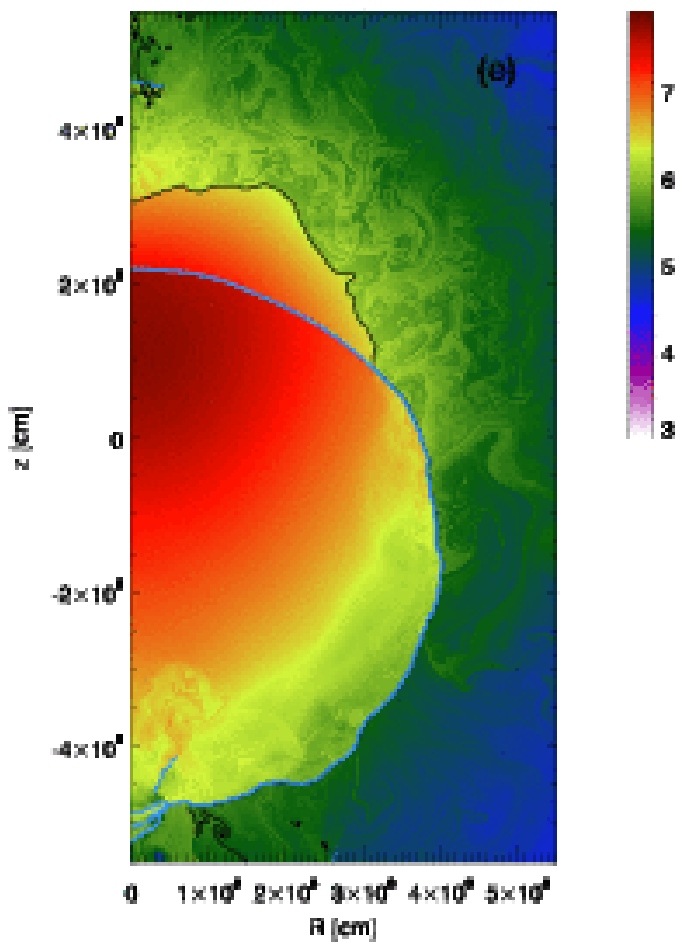}
\includegraphics[height=7.0cm,clip=true]{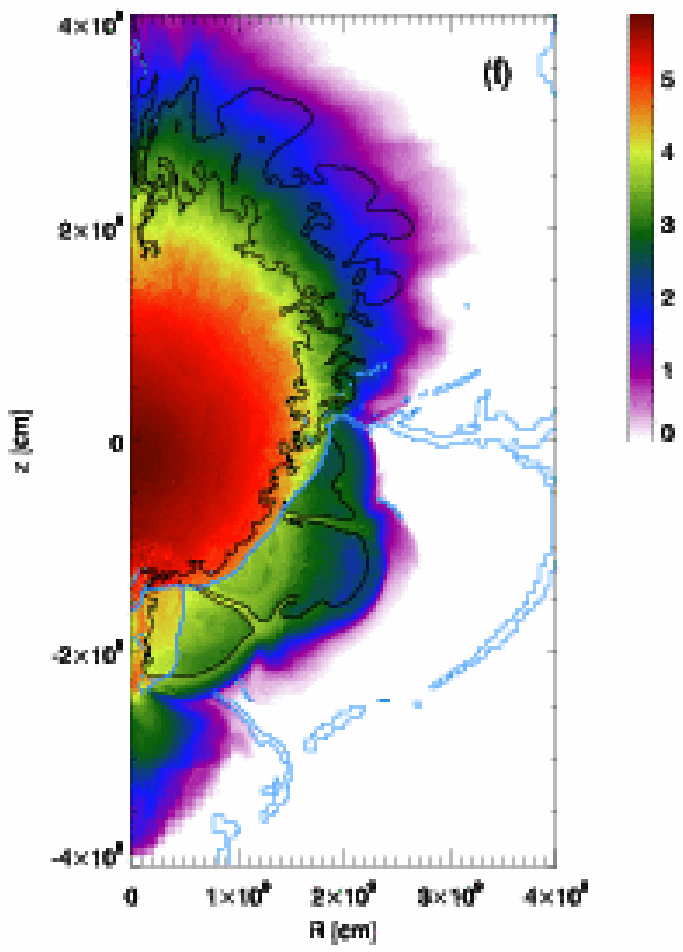}
\end{center}
\caption{Y12 model of a thermonuclear supernova explosion.  Shown is
the temporal evolution of density in log scale at times $t =
1.25,1.75,3.575,3.75,4.0,5.0$~s from panel (a) to (f) respectively
(note the change of spatial scale between panel (f) and the rest of
the sequence).  The position of the deflagrating material is marked
with black contours, while shockwaves are marked in light blue.  The
transition from deflagration to detonation is triggered by the shockwave
visible in panel (c) at $(r,z) = (0,-2.5 \times 10^8)$.  For more
details, see the text.}
\label{Fig:hydro}
\end{figure*}

\begin{figure}
\begin{center}
\includegraphics[width=8.0cm]{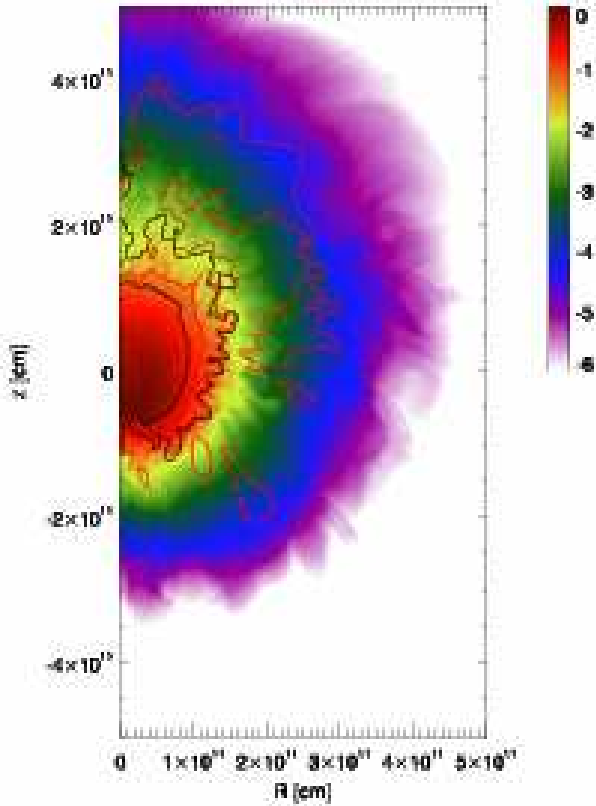}
\end{center}
\caption{Final density structure of the Y12 DFD model of a
thermonuclear supernova explosion in the homologous phase
 ($t=100$~s). The density is shown in log scale.  The spatial
coordinates can be converted to velocity space using $\vec{v}(r,t) =
10^{-7} \vec{r}(t)$~\kms.  The distributions of \nuc{Si}{28} and
\nuc{Ca}{40} are shown in red and black contours, respectively. The
contour levels correspond to $X(\nuc{Si}{28}) = 0.25$ and
$X(\nuc{Ca}{40}) = 0.01$.  The position of the inner IME-rich shell
can roughly be identified with the distribution of \nuc{Ca}{40}. The
material in the outer silicon-rich region was produced during the
deflagration phase.  The lack of calcium in the outer shell is the
result of the simplified nuclear burning method used to calculate the
deflagration.  Note the relative smoothness of the inner shell
compared to the inhomogeneity of the outer shell.}
\label{Fig:Homologous}
\end{figure}

\begin{figure}
\begin{center}
\includegraphics[width=8cm]{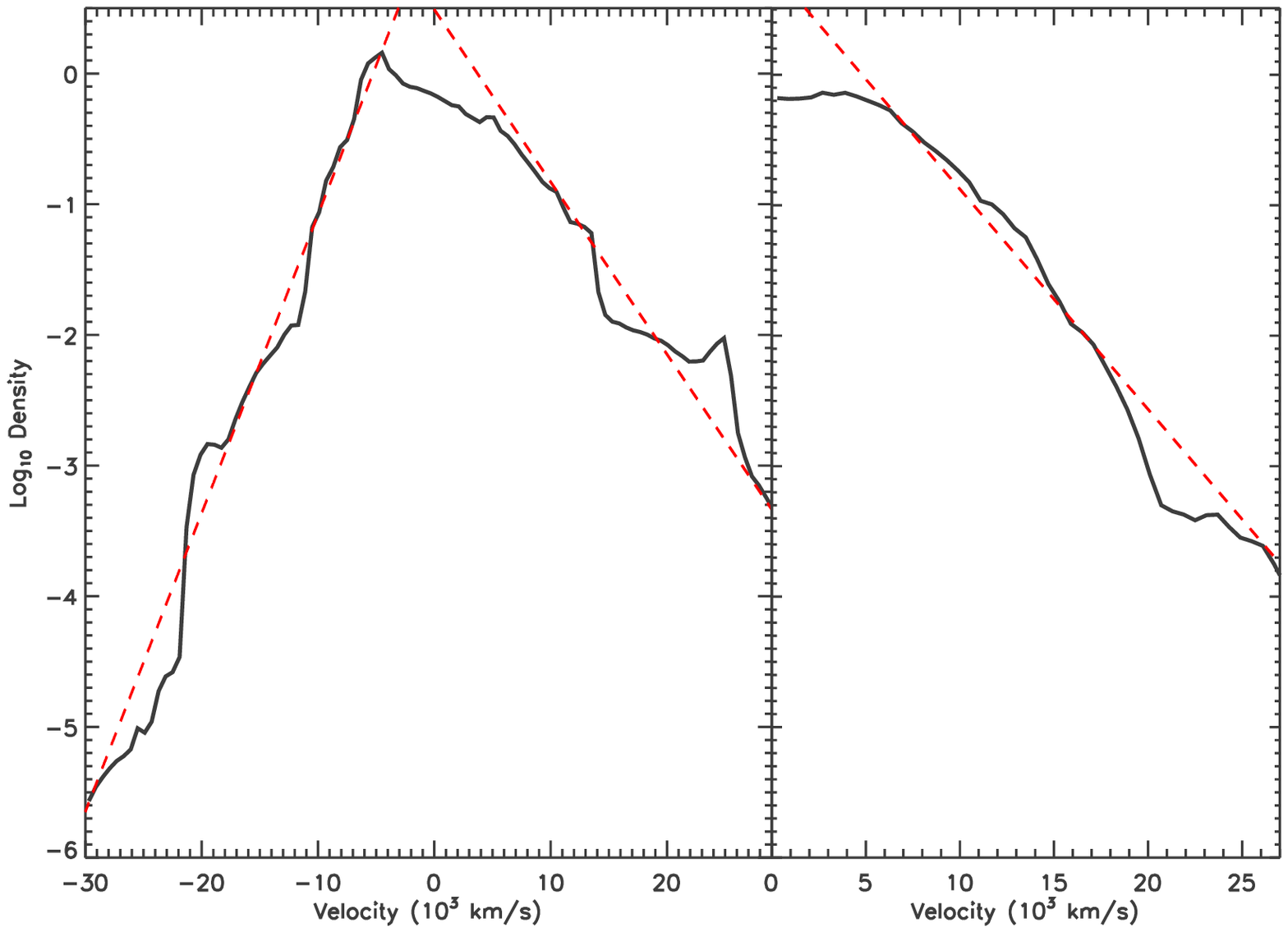}
\end{center}
\caption{Density structure (solid lines, in log scale) of the Y12 DFD model in the
homologous phase ($t = 100$~s) shown versus velocity coordinates.  The
dashed lines show the best fit exponential law $\rho(v) = \rho_0
\exp(-v/v_e)$. \emph{Left:} Density along the $z$-axis (the symmetry
axis) at $r=0$.  The ejecta is more extended for $z > 0$ (the ignition
side) and characterized by an exponential velocity scale $v_e =
3290~\kms$.  The ejecta is more compact for $z < 0$ (the detonation
side) with $v_e = 1900~\kms$.  The peak density is also slightly
offset from the expansion center.
\emph{Right} Density along the $r$-axis at $z=0$.  The density law is
characterized by $v_e = 2575~\kms$ }
\label{Fig:dens}
\end{figure}

The detonating failed deflagration (DFD) scenario of thermonuclear
supernovae considers an off-center, mild ignition process in a
degenerate Chandrasekhar mass carbon-oxygen white dwarf. In Paper~I,
we described detailed calculations of the explosion process as
performed with the multi-physics hydrodynamics code FLASH
\citep{Fryxell_flash}.  These calculations were all performed in two dimensions 
assuming axial symmetry.  From the set of the models presented in that
work, we have selected one example (Y12) for detailed analysis.  Here
we briefly review the salient features of the Y12 model from the point
of view of the calculation of model observables.

In the Y12 model, the white dwarf is initially ignited within a small
spherical region on the symmetry axis, 50~km in size and offset 12.5~km
from the center.  The resulting deflagration takes the form of a
buoyantly rising, high temperature bubble filled with freshly
synthesized nucleosynthetic products.  This early phase is illustrated
in Figure~\ref{Fig:hydro}a, in which the bubble surface (identified
with the edge of the deflagration front) is marked by the dark contour
line.

As the rising bubble approaches the white dwarf surface, the outer
layers of the star are deformed and laterally accelerated, leading to
the production of a circular surface wave.  By t=1.75~s
(Figure~\ref{Fig:hydro}b) that wave is seen to already have passed the
stellar equator.  By t=3.575~s (Figure~\ref{Fig:hydro}c) the wave
reconverges at the pole and begins radially compressing unprocessed
stellar material.  A jet-like flow develops in this confined region,
radially expelling material from the system while at the same time
forming a shock wave moving into the bulk of the star.  The shock-wave
is marked by the light blue line near $(r,z) \approx (0,-2.5\times
10^8$)~cm in Figure~\ref{Fig:hydro}c.

A short time later, the temperature in that shock-wave-dominated
region exceeds the ignition temperature for the carbon/oxygen mixture,
triggering a detonation through the shock-to-detonation mechanism
(Paper~I).  Unlike in previous delayed-detonation models
\citep[e.g.,][]{Hoeflich_DD, Gamezo_DD}, this transition to detonation is not 
inserted by hand but arises naturally in the course of the model
evolution.  Over the next 0.3~s, the detonation wave will
progressively consume the bulk of the star (Figures~\ref{Fig:hydro}d
and
\ref{Fig:hydro}e), and reprocess the material ejected in the initial
deflagration phase.  Burning ceases a few tenths of a second later.
By $t=5$~s (Figure~\ref{Fig:hydro}f), the supernova shockwave has
already completely swept through the portion of the star pre-expanded
by the deflagrating bubble $(z > 0)$, while it still remains trapped
in the denser portion of the star $(z < 0)$.

We followed the further evolution of the Y12 model up to $t=100$~s
after ignition.  At this time, the ratio of internal to kinetic energy
is found to be less than 1\% at all points in the ejecta, indicating
that the free expansion phase has been reached.  We verified that the
model velocity structure is homologous, that is, the local velocity is
linearly related to the radius $\vec{v}(r,t) =
\vec{r}(t)/t_0$ where $t_0 = 100$~s.  Thus
spatial coordinates can be transformed to velocity space using
$\vec{v}(r,t) = 10^{-7}~\vec{r}(t)$~\kms.  The total kinetic energy of
the explosion is $1.36\times 10^{51}$~ergs and the total mass of
\Nifs\ produced is 0.93~\msun.

The density structure of the model at this final time
(Figure~\ref{Fig:Homologous}) is nearly spherical, but with a slight
``egg-shape'', being more compact on one side.  As seen in
Figure~\ref{Fig:dens}, the characteristic ejection velocities are
higher on the side from which the bubble emerged (the ignition side)
and lower on the side from where the detonation took place (the
detonation side).  This is a result of the early expansion that
occurred preferentially on the ignition side due to the passage of the
deflagration bubble.

The compositional structure of the model, also illustrated in
Figure~\ref{Fig:Homologous}, is well stratified on the large scale,
and nearly spherical in shape.  Interestingly, the entire
compositional structure is offset by $\sim 3000$~\kms\ relative to the
center of expansion.  The innermost region of ejecta is rich in
iron-group elements (specifically \Nifs) and extends to $13,000$~\kms\
on the ignition side and $7000$~\kms\ on the detonation side.
Surrounding this is a shell rich in intermediate mass elements (IMEs),
spanning the velocity range $7000-10,000$~\kms\ on the detonation side
and $13,000-18,000$~\kms\ on the ignition side.  Both
\nuc{Si}{28} and
\nuc{Ca}{40} are present in this IME shell, as marked respectively by
the black and blue contour lines in Figure~\ref{Fig:Homologous}.
Nearly the entire star is consumed by the explosion, with very little
carbon/oxygen remaining.  

We note that the detailed nucleosynthesis calculated in the Y12 model
is only approximate (see Paper~I).  During the deflagration phase,
nucleosynthesis was calculated using an approximate burner which
tracked only the few most representative isotopes.  The detonation
phase employed a more extensive (but still approximate) 13-element
network. Post-processing of the model using a more complete nuclear
network is therefore necessary to more accurately specify the detailed
model abundances.  In particular, some fraction of the \Nifs\ produced
in the detonation phase may in fact represent stable iron group
elements, while some fraction of the silicon group produced in the
deflagration phase likely represents a distribution of various IMEs.

The inner layers of ejecta ($v \la 15,000~\kms$) in the Y12 model are
smooth and continuous, typical of the material processed solely by the
detonation.  This can be contrasted with the outermost layers of
ejecta, which were perturbed initially by the surface wave created by
the deflagration bubble.  That turbulent-like flow involved mixing of
the bubble material and unburned stellar matter, creating density and
compositional inhomogeneities with a typical size of about ten
degrees.

A characteristic feature of the model is the presence of two distinct
shells rich in IMEs.  The inner IME shell represents material
synthesized by the detonation wave.  The outer, high-velocity shell
($v \approx 25000-35000~\kms$) is composed chiefly of the material
created and expelled during the deflagration phase.  This
high-velocity shell resembles the pancake structure described in the
parameterized models of \cite{Kasen_GCD}, except that in the Y12 model
the material has a larger covering factor and a more irregular
structure.  Note that only silicon, and not calcium, is present in the
high-velocity shell, an artifact of the approximate burner used during
the deflagration phase.  One expects that, had a more complete nuclear
network been used, a significant amount of calcium would also be
present.  Rather than speculate as to that exact amount of calcium
produced, we leave the compositional structure as is, referring the
reader to the studies of \cite{Kasen_GCD}.

Before considering the observable properties of the Y12 model, we
mention a few limitations of the underlying explosion calculations.
Although no tuning of the model parameters was required to obtain an
explosion, the DFD mechanism itself is subject to uncertainties
regarding the initial conditions, geometrical simplifications, and
various numerical limitations.  In Paper~I, for example, a successful
shock to detonation transition (SDT) was not achieved for all ignition
configurations considered.  This may indicate a lack of robustness of
the underlying explosion mechanism, or may simply be a consequence of
the relatively low numerical resolution of the calculations. A low
resolution limits the size of the shocked regions, enhances numerical
diffusion, and as a result may underestimate the actual post-shock
temperatures and densities.  These factors may artificially decrease
the probability of SDT.  On the other hand, numerical diffusion is a
known source of mixing that may lead to artificial preheating and
spurious ignition of fuel \citep{fryxell+89}. Although the transitions
to detonation reported in Paper~I are due to shock compression in
nominally unburned material, the results are undoubtedly sensitive to
possible contamination due to mixing, which may affect the likelihood
of SDT.  Even if a SDT is not achieved, it is still possible for a
detonation to occur through the Zeldovich gradient mechanism
\citep{khokhlov_Zel} once the deflagrating material perturbs and mixes
with the surface layers of the star.

For completeness, we also summarize some more fundamental
simplifications of the DFD explosion model. In particular, we took the
progenitor to be a cold, static, non-rotating, unmagnetized,
chemically homogeneous white dwarf.  Although such a model is
considered "standard" in most modern SN~Ia explosion studies,
theoretical work strongly suggests that most of these assumptions are
in fact incorrect. The cores of massive white dwarfs are thought to be
both convective and chemically stratified
\citep{arnett69,hoeflich+02}.  Energy release from carbon burning
likely occurs even before ignition, which may contribute to the
pre-expansion of the progenitor and modify the energetics of the
subsequent deflagration.  Rotation of the white-dwarf can affect the
structure of the convective core and may induce a large scale
anisotropy of the explosion. Moreover, rotational shear may to some
extent change the evolution of the deflagration front and further
modify the effective flame energetics, which may influence the white
dwarf pre-expansion.  All of these elements necessarily contribute to the
previously mentioned numerical uncertainties of the DFD explosion
model, and deserve careful consideration in the future.

\section{Radiative Transfer Calculations}

The observable properties of the Y12 DFD model have been calculated
using the multi-dimensional time-dependent radiative transfer code
\Code\ \citep{Kasen_MC}. Given a homologously expanding SN ejecta
structure, \Code\ calculates the full time series of emergent spectra
at high wavelength resolution.  Broadband light curves are then
constructed by convolving the spectrum at each time with the
appropriate filter transmission profile. \Code\ includes a detailed
treatment of gamma-ray transfer to determine the instantaneous energy
deposition rate from radioactive \Nifs\ and
\Cofs\ decay.  Radiative heating and cooling rates are evaluated from
Monte Carlo estimators, and the temperature structure of the ejecta
determined by iterating the model to thermal equilibrium.

Several significant approximations are made in \Code, notably the
assumption of local thermodynamic equilibrium (LTE) in computing the
level populations.  In addition, bound-bound line transitions are
treated using the expansion opacity formalism (implying the Sobolev
approximation) and an approximate two-level atom approach to
wavelength redistribution.  Special care was taken for the calcium
lines, which were assumed to have pure scattering source functions for
the reasons discussed in \cite{Kasen_NIR}.  Note that
\Code\ allows for a direct Monte Carlo treatment of line fluorescence,
but due to computational constraints this functionality is not
exploited here.  See \cite{Kasen_MC} for a detailed code description
and verification, as well as tests of the expansion opacity and
two-level atom approximations.

In the atmospheres of SNe~Ia, the microscopic conditions assuring the
establishment of LTE ionization and excitation are in fact not met.
Nevertheless, a number of previous theoretical studies confirm the
adequacy of LTE models in reproducing the qualitative spectral and
photometric properties of SNe~Ia
\citep[e.g.,][]{Hoeflich_DD,Baron_NLTE, Pin01}.
The LTE simplification can be expected to result in quantitative
errors in the model light curves, especially at later times ($\sim
60$~days after explosion) when non-LTE ionization effects become
increasingly significant.  In addition, the model predictions in this
paper are subject to the further inadequacies of the two-level atom
framework and the available atomic line data \citep{Kasen_MC,
Kasen_NIR}.  We hope to reduce these uncertainties in future work by
relaxing some of the simplifying assumptions underlying the radiative
transfer calculations.

The time-dependent light curve calculations in this paper were
computed using the following numerical resolution: \emph{Spatial:} A
$100\times 50$ cylindrical grid extending to a maximum velocity of
30,000~\kms\ and with a cell length of 600~\kms; \emph{Temporal:} 100
time points spanning the epochs day~2 to day~80 with logarithmic
spacing $\Delta
\log t = 0.175$; \emph{Wavelength:} Spanning the range 100-25000~\AA\
with constant resolution of 100~\AA.  Extensive testing confirms the
adequacy of this resolution for the problem at hand.  A total of
$10^9$ photon packets were used in the Monte Carlo procedure.  The
emergent light curves and spectra were generated by collecting
escaping packets into 50 separate viewing angle bins.

The time-dependent \Code\ light curve calculation provides the full
spectral time series of the model, however to further improve our
synthetic spectrum results we performed further time-independent
calculations at five select epochs using both higher resolution
($200\times 100$ spatial grid, 5~\AA\ wavelength resolution) and a large
number ($10^8$) of packets.  The emergent bolometric luminosity
parameter for these static calculations was determined from the full
light curve calculation.  The static synthetic spectra achieve higher
signal-to-noise, but otherwise show excellent agreement with the
time-dependent spectra, except in small-scale features of size $<
100$~\AA.  In addition, one static spectrum near maximum light (\texp
= 18~days) was run using $10^{12}$~packets in order to study the $\la
1\%$ polarization levels.

\section{Comparison to Observations}\label{s:comparison}

We present below the synthetic optical/near-infrared light curves,
spectral time series, and spectropolarization of the Y12 model,
comparing and contrasting the results with observations of
representative SNe~Ia.  The properties of observed SNe~Ia vary
somewhat from object to object.  We make no attempt here to find an
optimal correspondence with the Y12 model, rather we simply compare to
the more well-known, well-studied objects.

\subsection{Light Curves and Color Evolution}
\label{Sec:LC}

\begin{figure*}
\begin{center}
\includegraphics[width=7in]{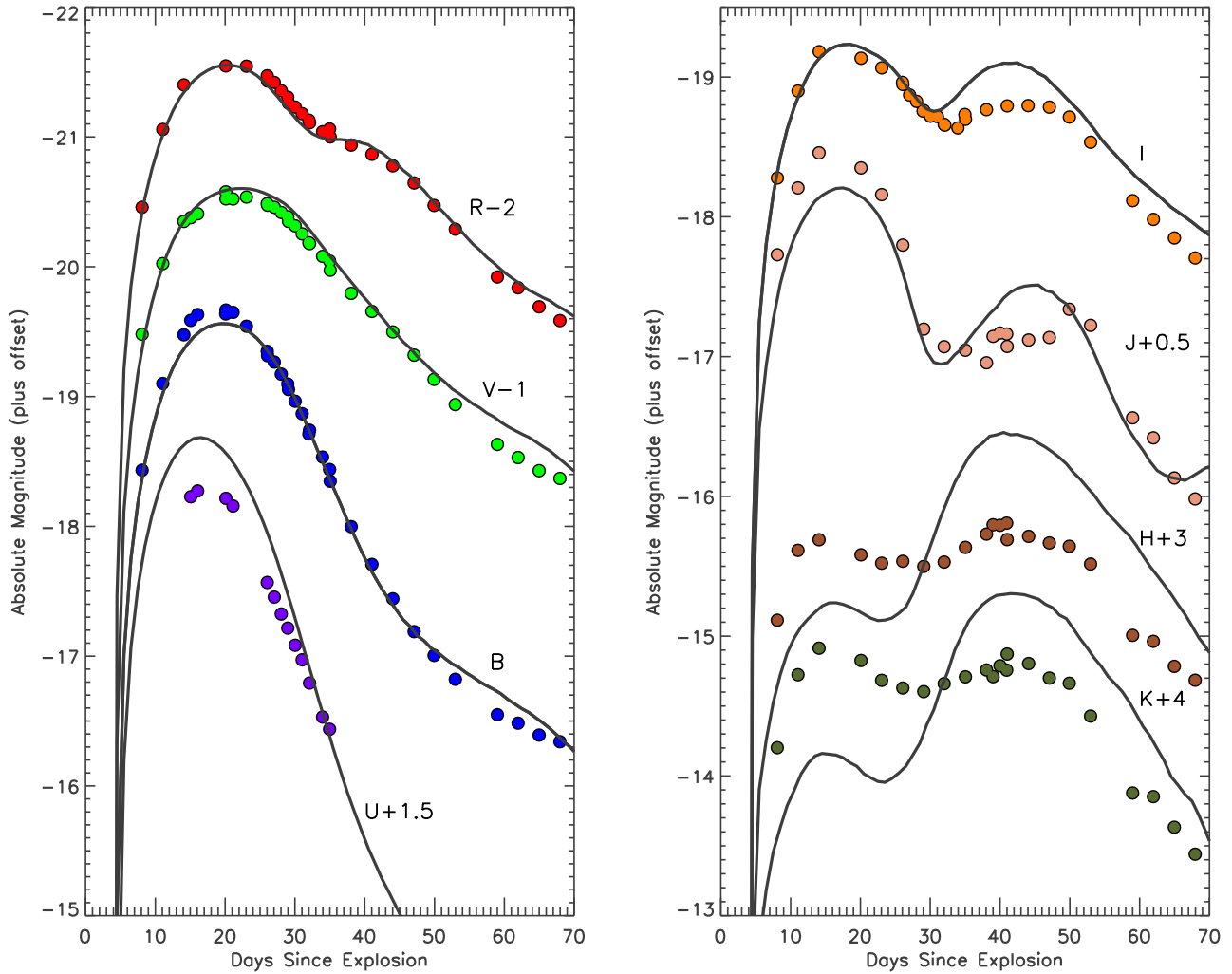}
\end{center}
\caption{Synthetic $UBVRIJHK$ light curves of the Y12 DFD model 
 as seen from an equatorial view ($\theta = 90^\circ$, solid lines)
 compared to observations of SN~2001el 
\citep[filled circles,][]{Kris_01el}. To align the models and observations, 
we adopt a distance modulus for SN~2001el of $\mu = 31.65$, which is
$0.35$~mag greater than the best estimate.}
\label{Fig:LC}
\end{figure*}

\begin{figure}
\begin{center}
\includegraphics[width=8cm]{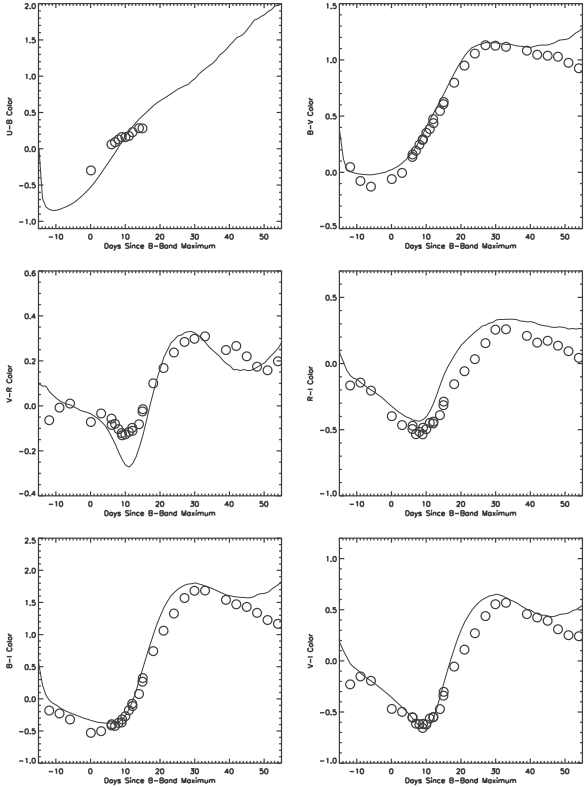}
\end{center}
\caption{Temporal evolution of the colors of the Y12 DFD model as seen from an equatorial view ($\theta = 90^\circ$, solid
lines) compared to observations of SN~2001el \citep[open
circles,][]{Kris_01el}. }
\label{Fig:Colors}
\end{figure}

\begin{figure}
\begin{center}
\includegraphics[width=8cm]{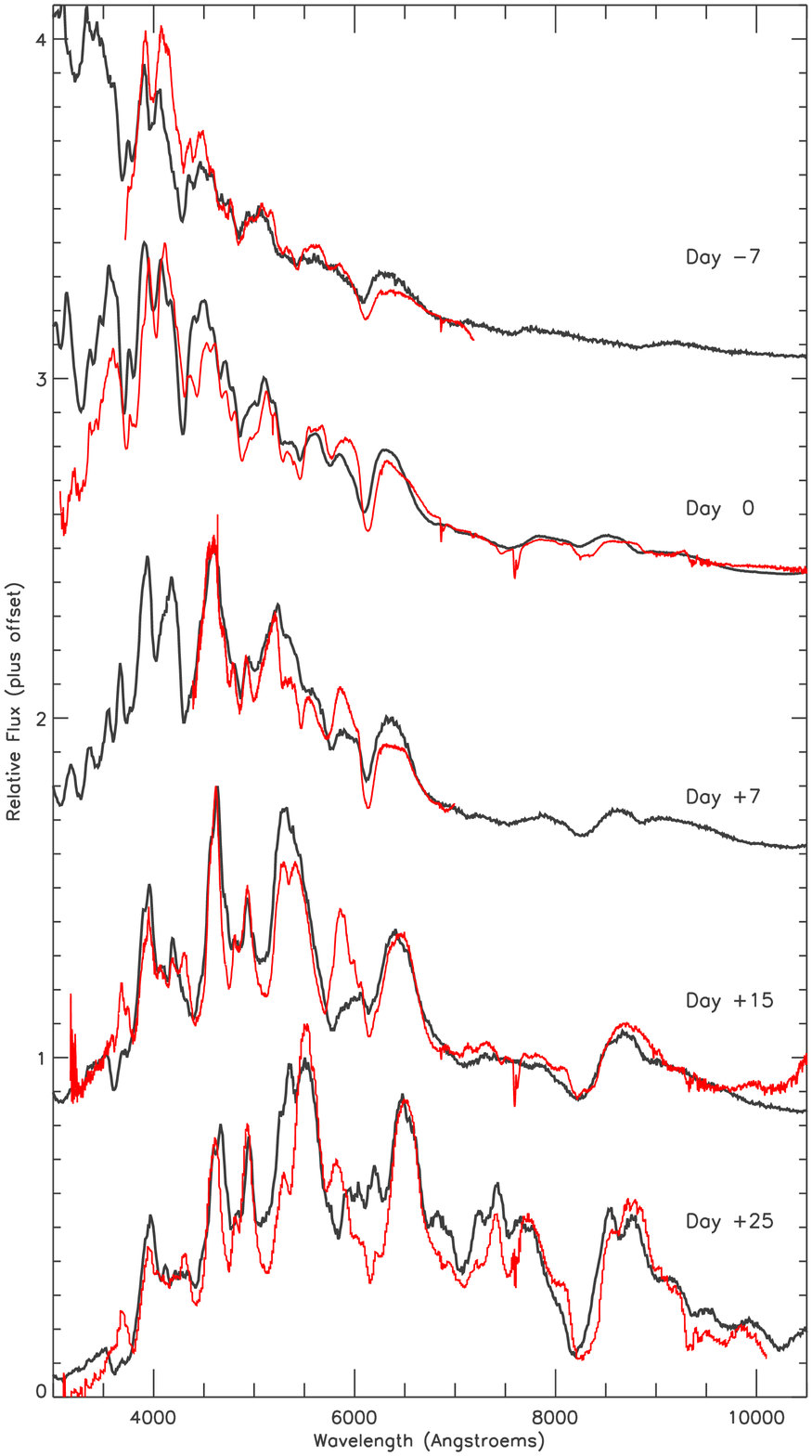}
\end{center}
\caption{Synthetic spectra at five different epochs 
of the Y12 DFD model as seen from an equatorial view ($\theta =
90^\circ$, black lines) compared to those of SN~1994D
\citep[red lines,][]{Patat_94D, Meikle_94D}. Labels give the day relative
to $B$-band maximum. }
\label{Fig:spectra}
\end{figure}

In Figure~\ref{Fig:LC}, we show the synthetic $UBVRIJHK$ light curves
of the Y12 model as compared to SN~2001el, a well observed and
photometrically normal SNe~Ia \citep{Kris_01el}. There is some
uncertainty in the total dust extinction to SN~2001el, which is
estimated between $A_v = 0.35-0.77$ depending on the method used.
Here we adopt the values $A_v = 0.57$ and $R_v = 2.88$
\citep{Kris_01el,Wang_01el} and correct for extinction using the dust
laws of \cite{CCM_dust}.

Given the intrinsic asymmetry of the Y12 model, we consider for now
only the equatorial view ($\theta = 90^\circ$) of the ejecta,
deferring discussion of the orientation effects to \S\ref{Sec:Orient}.
From this angle, the model light curves reach a peak absolute
magnitude of -19.57 in $B$, and -19.61 in $V$.  In Figure~\ref{Fig:LC}
we have adopted a distance modulus to SN~2001el of 31.65~mag so as to
align the observed peak magnitudes with those of the model.
\cite{Kris_01el} deduce a significantly lower distance of $31.29 \pm
0.08$~mag on the basis of the supernova photometry.  This suggests
that the Y12 model is $\sim 0.35$~mag too bright compared to SN~2001el
-- not surprising given the explosion produced over 0.9~\msun\ of
\Nifs.  This discrepancy may be mitigated somewhat if  detailed
nucleosynthesis calculations reveal that some fraction of the \Nifs\
in the Y12 model in fact corresponds to stable iron group elements.
Note also that the DFD scenario encompasses a family of explosion
models which produce greater or lesser amounts of \Nifs, depending on
the initial ignition conditions (Paper~I).

The shape of the model optical light curves match qualitatively those
of SN~2001el, especially in the $B$-, $V$- and $R$-bands.  The rise time
to $B$-band maximum is 18.8~days for the model, similar to that
observed among typical SNe~Ia \citep{Riess_Risetime, Conley_Risetime}.
The subsequent decline from peak is also quite consistent with the
observations of SN~2001el and other typical SNe~Ia, with a drop in
$B$-band magnitude 15 days after peak of $\dmfb = 1.19$~mag.  However,
given that brighter SNe~Ia generally display broader light curves, this decline
rate may be considered too fast in light of the high peak magnitude of
the model.  As we discuss in
\S\ref{Sec:Orient}, the decline rates of the model light curves in fact
depend rather sensitively on the orientation.

A distinct secondary maximum in the model $I$- and $J$-band light curves occurs
about three weeks after $B$-maximum.  A much weaker secondary maximum
is also present in the $R$-band.  The same double-peaked morphology is
seen in the SN~2001el observations, although the size and timing of
the secondary maximum differ somewhat from the model.  The properties
of the secondary maximum in fact vary substantially from supernova to
supernova
\citep{Nobili_Iband}, and also depend 
sensitively on the model parameters \citep{Kasen_NIR}.  Brighter
SNe~Ia (i.e., those with larger masses of \Nifs) typically display
more prominent secondary maxima.  The overly bright $I$-band secondary
maximum of the Y12 synthetic light curve indicates, in part, its overly large
mass of \Nifs\ compared to SN~2001el.

The model $H$- and $K$-band light curves show a much poorer
correspondence with the observations, with the secondary maximum being
much too bright relative to the first one.  As discussed by
\cite{Kasen_NIR}, this sort of discrepancy is attributable, in large
part, to the inadequacy of the atomic line data used in the transfer
calculations.  The Y12 model, however, shows a relatively poorer match
to SN~2001el when compared to other models considered in
\cite{Kasen_NIR}, likely due again to the relatively large
\Nifs\ mass.

The optical color curves of the model are compared to those of
SN~2001el in Figure~\ref{Fig:Colors}.  In general, the curves show
qualitative agreement up to about 40 days after $B$-band maximum.  At
later times, the model's color get progressively redder, while the
observations show the opposite trend.  It is at these late epochs,
when LTE predicts neutrality in the ejecta, that non-thermal
ionization by the products of radioactive decay becomes significant
\citep{Swartz_1991}.  Thus our transfer 
calculations should be considered less reliable at these phases.

\subsection{Spectral Evolution}
\label{Sec:Spectra}

The spectra of SNe~Ia consist of broad P~Cygni features superimposed
on pseudo-blackbody continuum.  By identifying individual absorption
features and measuring their Doppler-shifts, we probe the composition
and velocity of the absorbing ejecta material.  As the SN
material expands and geometrically dilutes, the photosphere recedes,
revealing emission from progressively deeper layers.  The spectral
time-series thus provides a scan of the multi-layered ejecta
compositional structure.

We choose to compare the synthetic spectra of the Y12 DFD model to
observations of SN~1994D, a well-sampled SN~Ia with normal spectral
characteristics \citep{Patat_94D, Meikle_94D}.  The dust extinction of
SN~1994D is believed to be very small, and we make no corrections for
reddening.
\cite{Branch_94D} provide a comprehensive line identification and 
spectroscopic analysis of SN~1994D.  In addition, SN~1994D has been
used to validate several other 1-dimensional explosion models such as
W7 \citep{Lentz_94D} and various delayed-detonation models
\citep{Hoeflich_94D, Baron_94D}, allowing cross-comparison of the Y12
model with other theoretical calculations.

In Figure~\ref{Fig:spectra}, we show the synthetic spectra of the Y12 model
(as seen from an equatorial view) at several different epochs -- viz.,
day $-7, 0, +7, +15,$ and $+25$ days with respect to $B$-band maximum.
In general, we find the synthetic spectra to be in impressively good
agreement with those of SN~1994D.  The quality of fit is comparable to
(and often better than) that obtained with classic 1-dimensional
models, such as W7 and the delayed-detonation models.

At maximum light ($\texp = 18.8$~days), the model reproduces all of
the major spectral features identified in the observations.  These are
primarily lines from IMEs, in particular three Si~II absorption
features (near 4000~\AA, 5850~\AA, and 6150~\AA), the S~II
``W''-feature (near 5200~\AA), and the Ca~II H\&K and IR-triplet
features (near 3600~\AA\ and 8200~\AA, respectively).  The only
feature not well fit at maximum light is the double absorption near
4600~\AA, due to blends of Mg~II and Si~III.  Interestingly, this
feature is peculiar in SN~1994D, and most other normal SNe~Ia more
closely resemble the model \cite[e.g., SN~1981B,][]{Branch_w7}.

The blueshifts of the absorption minima in the maximum light spectrum
indicate that the model velocities are in broad agreement with the
observations.  In detail, however, the model absorptions are
systematically bluer than the observations, most conspicuously in the
Si~II 6150 and Ca~II IR triplet lines.  We note that the blueshift of
the model absorption changes with orientation (see 
\S\ref{Sec:Orient}) and also is observed to vary from supernova to
supernova \citep{Benetti_Vel}.  The velocity differences noted in
Figure~\ref{Fig:spectra} are therefore within the observed scatter of
normal SNe~Ia.

A week after maximum light ($\texp = 25$~days), most of the same IME
features are still present in the observed spectra, while a prominent
Na~I feature emerges at 5600~\AA, blending with the smaller of the
Si~II features. The approximate alpha network used in the explosion
model did not include sodium, therefore the synthetic spectrum does
not reproduce the sodium feature.  This issue will be addressed in
future models by including more detailed nucleosynthesis.  Among the
other features, we note that the sulfur W-feature, well fit by the
model at maximum light, appears too weak at this epoch.

At later times, the SN photosphere recedes into the metal core of
ejecta, and the spectra progressively become dominated by lines from
iron group elements.  The model spectra at 14 and 21 days after
$B$-maximum well account for the observed features, with the exception
of the Na~I line.

\section{Intrinsic Model Asymmetry and Diversity}\label{s:diversity}
\label{Sec:Orient}

\begin{figure}
\begin{center}
\includegraphics[width=8.5cm]{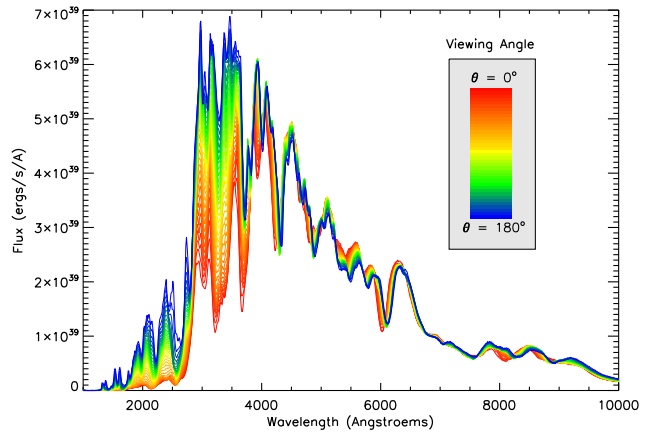}
\end{center}
\caption{Variation with viewing angle of the Y12 DFD model synthetic
spectrum at maximum light (\texp = 18~days). The angle $\theta =
0^\circ$ corresponds to the view from the ignition side ($z > 0$ in
Figure~\ref{Fig:Homologous}). }
\label{Fig:spec_los}
\end{figure}

\begin{figure}
\begin{center}
\includegraphics[width=8.5cm]{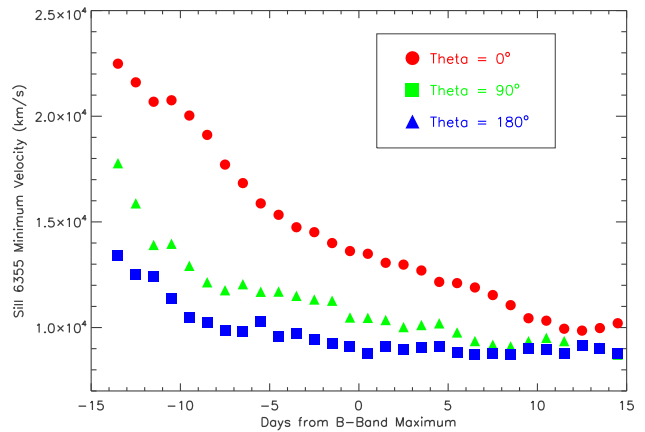}
\end{center}
\caption{Temporal evolution of the Si~II 6150 absorption minimum
velocity as seen from three different viewing angles. The angle
$\theta = 0^\circ$ corresponds to the view from the ignition side ($z
> 0$ in Figure~\ref{Fig:Homologous}). while $\theta = 90^\circ$
corresponds to an equatorial view and $\theta = 180^\circ$ corresponds
to the view from the detonation side ($z <0$). The variation in the
model velocity evolution can be compared the observed diversity in
SNe~Ia \citep[][Figure~1] {Benetti_Vel}.}
\label{Fig:si_vel}
\end{figure}

A major advance offered by our multi-dimensional radiative transfer
calculations is the ability to study the dependence of the model
observables on the viewing angle.  In the DFD model, the fundamental
orientation effects arise from the variation of the characteristic
ejecta expansion velocities along different directions.  As discussed
in \S\ref{s:y12} (see also Paper~I) this distortion reflects the
inherent asymmetry of the initial conditions and subsequent
deflagration.  The global asymmetry of the partially burned white
dwarf is then imprinted on the dynamical and compositional structure
of the ejecta by the off-center detonation.

\subsection{Photometric and Spectroscopic Diversity}
\label{Sec:specdiv}

The asymmetry in the Y12 ejecta structure leads to two significant
orientation effects in the model spectrum, both clearly visible in the
maximum-light spectrum shown in Figure~\ref{Fig:spec_los}.

First, the blueshift of most spectral absorption features depends on
the viewing angle, becoming larger as one looks closer to the ignition
side.  This is due to the generally higher expansion velocities of the
IME shell on that side of the ejecta.  Figure~\ref{Fig:si_vel}
quantifies this Doppler shift dependence by plotting the time
evolution of the velocity of the prominent Si~II absorption near
6150~\AA, as seen from several viewing angles.  In this figure we have
measured the velocity directly from the Doppler shift of the Si~II
absorption minimum for each of the synthetic spectra in our daily
spectral time-series.  Such a method corresponds exactly to what is
done in the observational studies.  By contrast, most other
time-dependent SN radiative transfer calculations determine only the
velocity of the SN photosphere, which may or may not coincide with
that of the Si~II line.

Near maximum light, the Si~II velocity varies from $9,000$ to
$13,500$~\kms, or by nearly 40\%, depending on the viewing angle.  The
velocity is largest and has the fastest rate of decrease with time
when observed near the ignition side ($\theta = 0^\circ$).  For
viewing angles away from the ignition side, the velocity of the Si~II
feature is lower and almost constant with time.  This variety of
velocity gradients can be compared to Figure~1 of \cite{Benetti_Vel},
which demonstrates a strikingly similar diversity among observed
SNe~Ia.  The Y12 model prediction is therefore, at first glance, in
reasonable agreement with the statistical behavior of the
observational sample. On the other hand, the cluster analysis of
\cite{Benetti_Vel} suggests that observed SNe~Ia can be divided into two
distinct classes of ``high'' and ``low'' velocity gradient.  In the
Y12 model, by contrast, the orientation effect leads to a range of
events intermediate  the two classes.  This continuous behavior
appears, at first glance, inconsistent with the dichotomy of
\cite{Benetti_Vel}, although a detailed statistical comparison would
be needed to truly confirm the discrepancy.

The second consequence of the model asymmetry on the maximum light
spectrum is that the flux level at shorter wavelength shows a strong
variation with viewing angle.  In the $B$-band ($4000$~\AA~$\la
\lambda \la 5000$~\AA) this variation is only of order $10\%$, however
it increases to a factor of $\sim 4$ in the $U$-band ($3200$~\AA~$\la
\lambda \la 4000$~\AA) and is even larger in the
ultraviolet.  The effect is a result of the aspherical distribution of
iron group elements in the ejecta. On the ignition side ($\theta =
0^\circ$) iron group species extend to a velocity $v
\approx 13,000$~\kms, whereas on the detonation side ($\theta =
180^\circ$) these elements are restricted to velocities $v \la
7,000$~\kms.  It is primarily the line blanketing from iron group
lines that determines the flux level at shorter wavelengths.  When
viewed at $\theta = 180^\circ$, the relative lack of iron group
elements at higher velocities leads to significant reduction in the
degree of line blanketing, and hence a spectrum that is brighter at
bluer wavelengths.

\begin{figure}
\begin{center}
\includegraphics[width=8.5cm]{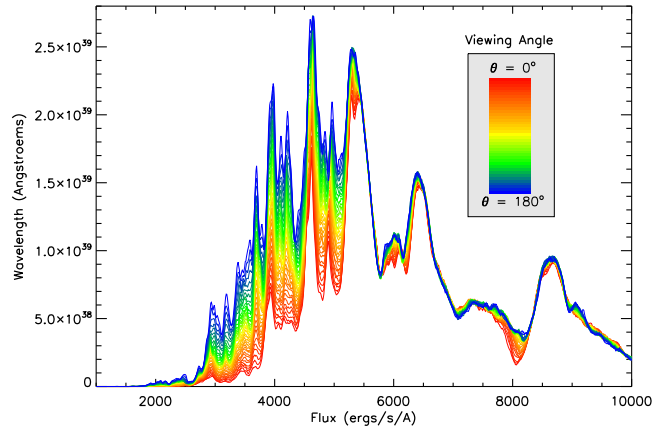}
\end{center}
\caption{ Variation with viewing angle of the Y12 DFD model synthetic
spectrum two weeks past maximum light (\texp = 32~days). The angle
$\theta = 0^\circ$ corresponds to the view from the ignition side ($z >
0$ in Figure~\ref{Fig:Homologous}) Note that the strong flux variation
has shifted redward when compared to that at maximum light
(Figure~\ref{Fig:spec_los}). }
\label{Fig:spec_los_d32}
\end{figure}

\begin{figure}
\begin{center}
\includegraphics[width=8.5cm]{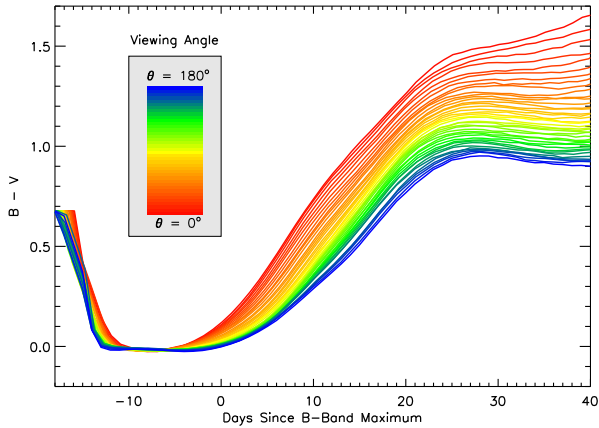}
\end{center}
\caption{ Variation with viewing angle of the Y12 DFD model synthetic
$B-V$ color with time. The angle $\theta = 0^\circ$ corresponds to the
view from the ignition side ($z > 0$ in
Figure~\ref{Fig:Homologous}). }
\label{Fig:col_los}
\end{figure}

At two weeks after maximum light, the same sort of orientation effects
are visible in the synthetic spectra (Figure~\ref{Fig:spec_los_d32}).
By this epoch, the layers of ejecta rich in iron-group elements have
begun to recombine from doubly to singly ionized, and features from
Fe~II/Co~II lines dominate the spectrum.  Again, because of the
aspherical distribution of iron group elements, the flux level at
shorter wavelengths shows a strong dependence on the viewing
angle. However, at this time the asymmetry affects wavelengths even
longer than at maximum light, as Fe~II/Co~II lines have become
prominent in the optical part of the spectrum.  The $B$-band portion
of the spectrum now shows flux variations by a factor of nearly two.
This line-blanketing effect leads to a strong dependence of the model
$B-V$ color evolution on orientation, especially after maximum light,
as seen in Figure~\ref{Fig:col_los}.

\begin{figure*}
\begin{center}
\includegraphics[width=6.0in]{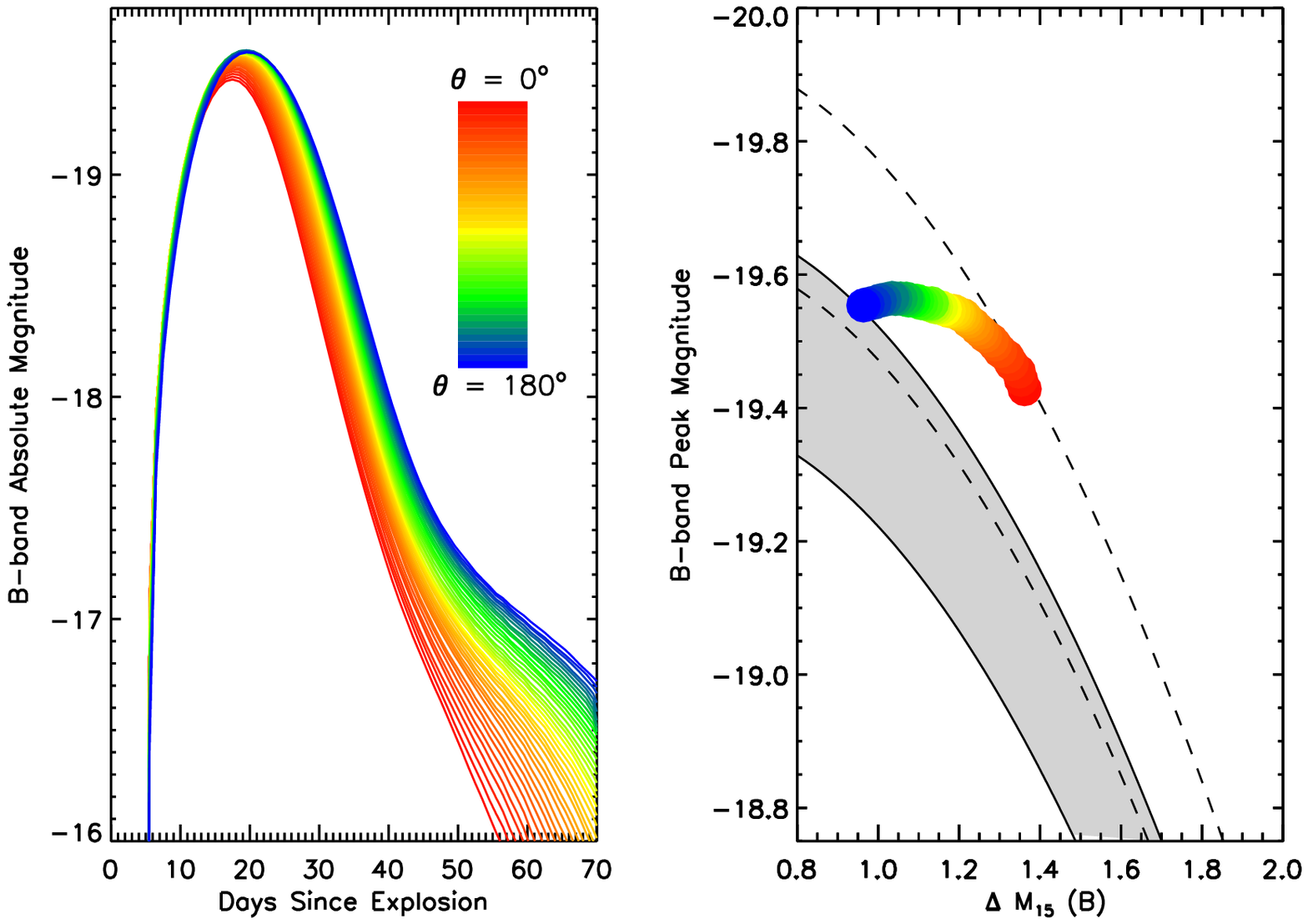}
\end{center}
\caption{\emph{Left:} Variation of the Y12 model $B$-band
light curve with viewing angle. $\theta = 0$ corresponds to the view
from the ignition side ($z > 0$ in
Figure~\ref{Fig:Homologous}). \emph{Right:} Relationship between the
model peak $B$-band magnitude $M_B$ and decline rate \dmfb\ as seen
from different viewing angles.  Color coding is the same as in the
left panel.  The shaded region is the observed relation of
\cite{Phillips_1999} with a calibration $M_B = -19.3$~mag at $\dmfb =
1.1$~mag and spread of $\sigma = 0.15$~mag.  The dotted lines show the
same relationship shifted upward by $0.25$~mag. }
\label{Fig:BLC_los}
\end{figure*}

Clearly this orientation effect has important implications for the
$B$-band light curves of the model, and consequently the SN~Ia
width-luminosity relationship.  The development of Fe~II/Co~II line
blanketing largely determines the $B$-band decline rate \dmfb\ of
SNe~Ia \citep{KW06}.  As seen in Figure~\ref{Fig:BLC_los}, the
aspherical distribution of iron group elements in the Y12 model thus
leads to significant variations in
\dmfb\ with viewing angle.  The decline rate changes from
\dmfb = 0.96~mag at $\theta = 0^\circ$ to
\dmfb = 1.36~mag at $\theta = 180^\circ$.  This is a good fraction
of the total range of decline rates noted in the observational sample
of SNe~Ia.

The peak $B$-band magnitude of the Y12 model also varies with viewing
angle, although to a lesser extent, from $M_B = -19.43$~mag at $\theta
= 0^\circ$ to $M_B = -19.56$~mag from the opposite side.  The
correlation between \dmfb\ and $M_B$ thus has the same sign as the
observed width-luminosity relation, in that broader SNe are generally
brighter.  However, as seen in Figure~\ref{Fig:BLC_los} (right panel),
the correlation is too weak as compared to the observed relation of
\cite{Phillips_1999}.  Thus the asymmetry of the Y12 model could at
most be considered as a potential source of dispersion in the
width-luminosity relation.  The model intrinsic dispersion is $\sigma
\approx 0.13$~mag, roughly equal to the level noted in
observations \citep{Hamuy_96b} and leaving little room for a
contribution from observational or other errors.

As also seen in Figure~\ref{Fig:BLC_los} (right panel), the light
curves of the Y12 model appear to be systematically brighter than the
observations, given their range of \dmfb.  The shaded band in the
figure is the width-luminosity relation of \cite{Phillips_1999} with
an assumed calibration of $M_B = -19.3$~mag for $\dmfb = 1.1$~mag and
with a dispersion $\sigma = 0.15$~mag.  The model points lie roughly
$0.25$~mag above this relation.  The significance of this discrepancy
is debatable, as the absolute calibration of the empirical
width-luminosity relation is subject to uncertainties in the distances
to SNe~Ia, while the model light curves may possess systematic errors
of order 0.1-0.2~mag due to approximations made in the radiative
transfer calculations.

The variation in the $B$-band magnitude with viewing angle becomes
even more dramatic at later times, and exceeds 1.5~mag at 40 days
after $B$-maximum.  This is larger than the $\sim1$~mag dispersion
seen in the observational sample \citep{Riess_22LC}.  Note, however,
that the increased uncertainties in the radiative transfer at later
times (as noted in \S\ref{Sec:LC}) may contribute to the extreme spread in
model $B$-magnitudes at these epochs.

\subsection{Polarization}

\begin{figure}
\begin{center}
\includegraphics[width=8.5cm]{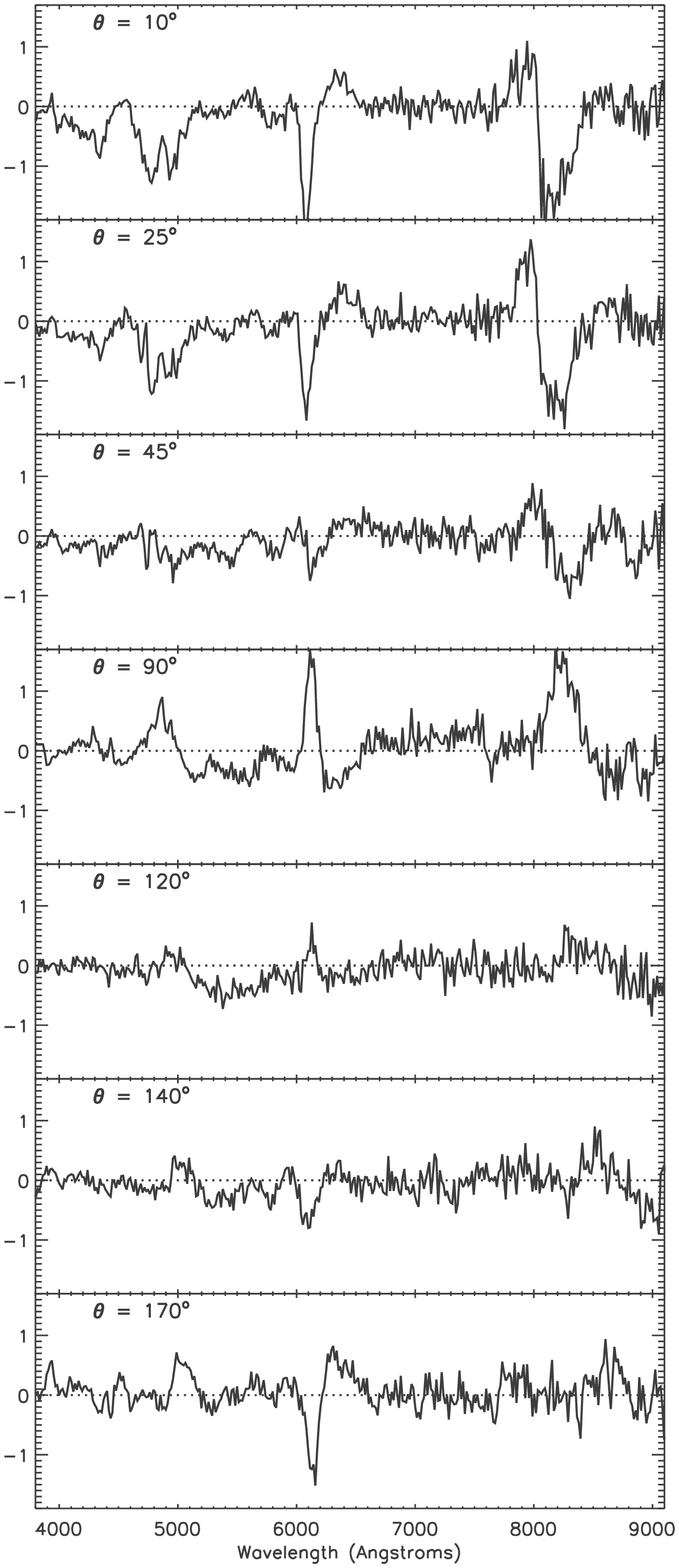}
\end{center}
\caption{Synthetic polarization spectrum of the Y12 DFD model near
maximum light (\texp = 18~days) as seen from several viewing angles
$\theta$.  Small scale fluctuations represent the level of Monte Carlo
noise in the calculation.}
\label{Fig:polspec}
\end{figure}

\begin{figure}
\begin{center}
\includegraphics[width=8.5cm]{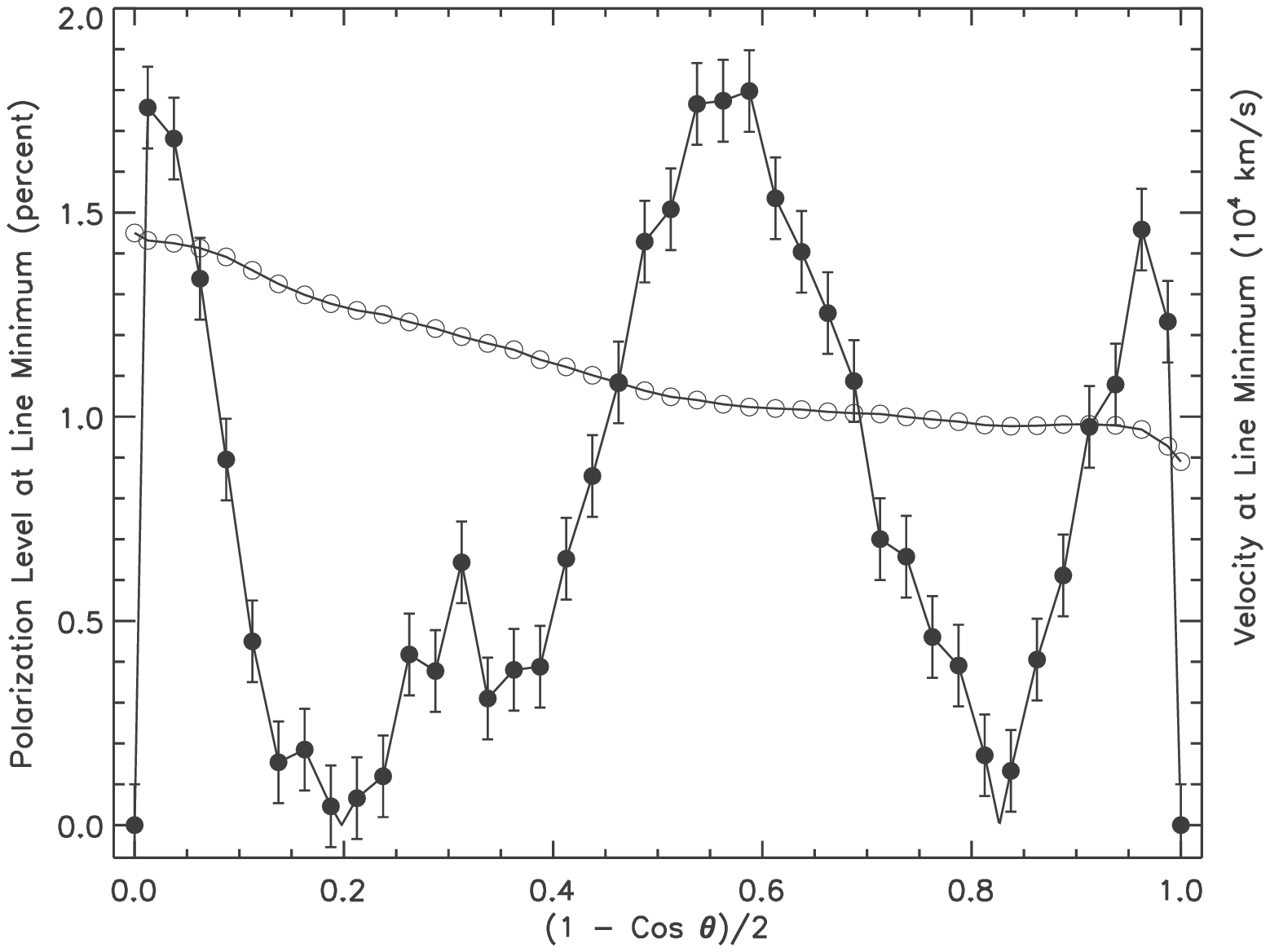}
\end{center}
\caption{Velocity (open circles, in $10^4$~\kms) and 
absolute value of the polarization (closed circles, in percent)
measured at the line minimum of the Si~II 6150 feature of the Y12
model at maximum light.  The size of the error bars ($0.1\%$) roughly
represents the level of Monte Carlo noise in the polarization
calculations.  The abscissa is $x = (1 -
\cos\theta)/2$ where $\theta$ is the viewing angle. The
probability of observing the model within the range $x = (x_1,x_2)$ is
then easily figured as $p = x_2 - x_1$.  The polarization cancels by
symmetry at $\theta = 0^\circ$ and $180^\circ$ and so has been set to
zero at these points.}
\label{Fig:vpol}
\end{figure}

Polarization measurements provide the most direct probe of the
asphericity of the SN ejecta.  Light is polarized by electron
scattering in the hot, ionized SN envelope.  Detection of a finite net
polarization indicates a preferred direction in the scaterring medium,
and hence a departure from spherical symmetry.  Spectropolarimetry is
a challenging endeavor, as the light from SNe is dim and quickly
fading, and the observed polarization levels are typically small, $P
\la 1\%$.  Nevertheless, intrinsic polarization has now been clearly 
detected in several SNe~Ia, both in the lines and the continuum
\citep{Howell_99by,Wang_01el,Leonard_SNIa,Chornock_05hk}.  The sample
size of objects with high signal-to-noise spectropolarimetry is
unfortunately still rather small, and the observations are also
subject to uncertainties in the degree of interstellar polarization
caused by scattering by dust grains along the line of sight.

The continuum polarization level (i.e., the degree of polarization at
wavelengths where electron scattering is the dominate opacity)
provides the most basic measure of global asymmetry in the bulk SN
ejecta.  For the Y12 model, the continuum polarization is in fact very
small during the epochs around and after maximum light.  This reflects
the fact that the density contours of the ejecta are not grossly
aspherical.  Moreover, the primary mode of the model's ``egg-shaped''
density structure is on a $180^\circ$ scale, whereas the continuum
polarization measures differences oriented $90^\circ$ apart.  In
particular, while the polarization of light scattered from the
ignition side of the ejecta is enhanced compared to a spherical model,
this is nearly counter-balanced by the relative decrease in the
polarization of light scattered from the compact detonation side
$180^\circ$ away.  The polarized flux from the polar regions of the
Y12 model then nearly equals the orthogonally oriented polarized flux
from the equatorial regions.  We therefore find that at maximum light,
the continuum polarization of the Y12 model is small, $P < 0.1\%$,
from all viewing angles.  This is smaller than the continuum
polarization of $P
\approx 0.3\%$ typically observed in SNe~Ia
\citep[e.g., SN~2001el,][]{Wang_01el}.  Given the signal-to-noise of
most spectropolarimetry observations, a polarization of $P < 0.1\%$ is
usually indistinguishable from zero.

Although the continuum polarization of the Y12 model is small at
maximum light, the polarization over line features can be
substantial. In Figure~\ref{Fig:polspec} we show the synthetic model
polarization spectrum at maximum light, as seen from several viewing
angles.  Although nearly $10^{12}$ photon packets were used in this
Monte Carlo calculation, the synthetic polarization spectra still
show random noise at a level $\sim 0.1$\%.  Nevertheless, the broad
line features are easily discernible among the high-frequency
fluctuations.  The most prominent polarization features are due to the
Si~II absorption feature near 6150~\AA, the Ca~II IR triplet near
8200~\AA, and the blend of Fe~II lines near 4900~\AA.  The
polarization over the Si~II absorption feature reaches nearly 2\% from
certain viewing angles.

The line polarization of the Y12 model reflects the global asymmetry
of the ejecta compositional structure, which results in an
asymmetrical distribution of the line opacity.  From certain viewing
angles, the opacity from a strong line partially eclipses the
underlying electron scattering photosphere, leading to an incomplete
cancellation of the polarization over the wavelengths of the line
absorption feature
\citep{Kasen_01el}.  In the Y12 model, the most important 
asymmetry is that the distribution of IME-rich material extends to a
higher velocity on the ignition side of the ejecta.  In some respects,
this asymmetry resembles that of the ``pancake'' geometry studied in
\cite{Kasen_GCD}. although in the present case the excess of IME-rich
material abuts the primary IME shell, rather than being detached from
it. The dependence of the line polarization on viewing angle thus
resembles that explained in \cite{Kasen_GCD} -- in particular, the
line polarization is negative for small viewing angles ($\theta \la
45^\circ$) but flips sign for intermediate viewing angles ($\theta
\approx 90^\circ$).

Although we have no direct way to determine the viewing angle itself
from SN observations, what we can study is the correlations between
different angle dependent observables.  In particular, it is useful to
study the relation between the velocity of a line and its
polarization, where both quantities are measured at the minimum of the
flux absorption feature.  In Figure~\ref{Fig:vpol}, we plot these
values for the Si~II 6150 feature of the Y12 model, as derived from
the maximum light polarization spectrum.  In the figure, we chose the
abscissa to be $x = (1 - \cos\theta)/2$, where $\theta$ is the viewing
angle.  The probability of observing the model within the range $x =
(x_1, x_2)$ is then simply given by $p = x_2 - x_1$.

As discussed already (\S\ref{Sec:specdiv}), for viewing angles towards
the ignition side ($x \la 0.1$ or $\theta \la 35^\circ$) the Si~II
velocity is unusually high ($v \approx 14,000$~\kms).  From these
angles the Si~II polarization is also always large ($P
\sim 1-2\%$).  A similar trend appears in the observations.  
Among the limited number of SNe~Ia with spectropolarimetry, the Si~II
polarization is found to be largest in the two objects with the
highest Si~II absorption velocities
\citep[SN~2002bf and SN~2004dt,][]{Leonard_SNIa}.  The Y12 model
predicts such events should be observed roughly 10\% of the time.

From viewing angles away from the ignition side, the Si~II
polarization of the Y12 model may be either small or large.  For
$45^\circ \ga \theta \ga 78^\circ$ ($x = 0.15-0.4$), the model
generally reproduces the low polarization levels and moderate line
velocities characteristic of typical events like SN~2001el
\citep{Wang_01el} and SN~2003du \citep{Leonard_SNIa}.  However, the
Y12 model also predicts that some SNe~Ia should exhibit high Si~II
polarization accompanied by low or moderate Si~II line velocity.  In
particular, for $x = (0.45,0.70)$ and $x = (0.9,1.0)$ the polarization
is $P \la 1.0\%$ while the velocity is only $v \approx
9,000-10,000$~\kms.  According to the model, such objects should be
observed roughly 35\% of the time, however no SN~Ia with this
characteristic has yet been found.

On the whole, the Y12 model predicts that the Si~II line polarization
will be greater than 1\% from almost half of all possible viewing
angles, which is generally higher than the average level of line
polarization noted in the current (albeit limited) observational
sample of SNe~Ia.  This suggests that the degree of compositional
asymmetry in the model ejecta structure may be too extreme.  This same
conclusion had already been indicated by the large dispersion in the
$B$-band decline rates (\S\ref{Sec:specdiv}).

In addition to a global asymmetry, the IME layer of the Y12 model
possesses numerous compositional inhomogeneities (``clumps'') of
typical size ten degrees.  These were the result of the original
turbulent-like flow of the deflagration material across the white
dwarf surface.  The signatures of these clumps, however, are not
readily discernible in the polarization spectra of
Figure~\ref{Fig:polspec}.  In this 2D calculation, the global
asymmetry of the IME distribution dominates the polarization signal.
In a fully 3D calculation, on the other hand, the IME clumps would
lead to some deviation of the line polarization from a single axis of
symmetry.  This spectropolarimetric signature has been used to infer
clumpiness in the IME layers of some observed SNe~Ia
\citep{Wang_04dt}.  In this sense, the DFD scenario is notable in its
ability to produce a clumpy outer layer of IME while retaining a
rather smooth distribution in the inner ejecta.

The Y12 model also possesses clumps of IME at very high velocity ($v =
25,000-35,000$~\kms) representing material produced and expelled
during the deflagration phase.  Assuming these clumps were to possess
a moderate abundance of calcium, one would expect to observe high
velocity, highly polarized Ca~II IR triplet absorption features in the
maximum-light spectrum, similar to those discussed in
\cite{Kasen_GCD}.  However, because the approximate nucleosynthesis
employed during the deflagration phase of the explosion calculation
did not include calcium, the high-velocity material does not in fact
influence our current synthetic flux or polarization spectra.  Given
the low densities in these layers, the lines from all other elements
besides calcium are too weak to be seen.

More detailed and conclusive studies of the ejecta geometry of SNe~Ia
will be possible once the database of spectropolarimetry observations
is expanded, such that for each subclass of SNe~Ia we have enough
events to sample all viewing angles in a uniform way (e.g., always at
exactly the same epoch).  Figure~\ref{Fig:vpol} is just one
demonstration of how we can use such a data set to powerfully
constrain multi-dimensional models of SNe~Ia.

\section{Summary and Conclusions}

We presented the model observable properties of one example of the DFD
explosion scenario of thermonuclear supernovae.  The explosion model
involved the off-center ignition of Chandrasekhar-mass white dwarf
which later underwent a transition from deflagration to detonation and
produced an energetic explosion.  The model was expanded well into the
homologous phase, and then used as input to time-dependent
multi-dimensional radiative transfer calculations.  The transfer
calculations supplied the emergent broadband light curves, color
evolution, spectral time-series, and polarization of the model, all of
which were critically evaluated against examples of well-observed,
standard SNe~Ia.  

The calculations presented here were intended primarily to develop and
apply a methodology of validating multi-dimensional SN models
against observations.  They represent the first extensive comparison
of a self-consistent multi-dimensional SN~Ia model against
observations, and comprise one of the most comprehensive evaluations
of any SN model to date.  On the whole, the properties of the Y12 DFD
model appear fairly consistent with observations, although certain
discrepancies are also apparent.  We summarize the primary successes
and failures of the model as follows:
\begin{itemize}
\item The shape of the model broadband light curves resembles
observations of the normal Type~Ia SN~2001el.  However, given the
model's large \Nifs\ mass (0.93~\msun), the synthetic light curves
are generally brighter than the typical SNe~Ia, by about 0.35~mag.  The
model light curves reach a peak absolute magnitude of -19.57 in $B$,
and -19.61 in $V$, as viewed from the equatorial direction ($\theta =
90^\circ$).
\item The $B$-band decline rate of the model ($\dmfb = 1.19$~mag, 
as viewed from the equatorial direction) is typical of an average
SN~Ia, but too fast when compared to observed SNe~Ia of equivalently
high peak brightness.
\item The $R$-, $I$-, and $J$-band light curves of the model exhibit a secondary maximum 
which resembles qualitatively that seen in observations.  The timing
and prominence of the model secondary maximum differs somewhat from
that seen in SN~2001el, likely due to the model's relatively large
\Nifs\ mass.
\item The $B-V$ and $V-R$ color curves of the model (from $\theta = 90^\circ$)
fit those of SN~2001el to within $\sim0.1$~mag up to day 30 after
maximum light.  The $U$- and $I$-band light curves more strongly
deviate from the observations, leading to a slightly poorer, but still
reasonable match between the model $U-B$, $B-I$, $V-I$, and $R-I$
color curves and observations.
\item Inspection of the model spectra (from an equatorial view) each
week from -7 to +25 days relative $B$-maximum shows good agreement
with observations of the spectroscopically normal SN~1994D.  The
quality of fit is comparable to that of standard 1D explosion models
such as W7 and the 1D delayed-detonations.
\item The asymmetry of the model ejecta structure
leads to intriguing orientation effects.  The blueshifts of individual
absorption features vary with viewing angle by more than 40\% at
maximum light.  Moreover, depending on the viewing angle, the
evolution of the expansion velocities (as measured from the minimum of
the Si~II 6150 absorption) accommodate the entire range of observed
behaviors, from high to low velocity gradient
\citep{Benetti_Vel}.
\item Due to the asymmetrical velocity distribution of iron group elements,
first the ultraviolet and later the blue part of the spectrum show
substantial variation (by a factor of several) in flux absorption
along different viewing angles. This leads to a significant dependence
of the $B$-band decline rate on orientation, with \dmf\ varying between
$0.96-1.36$~mag.  The model shows a correlation of \dmfb\ with the
$B$-band peak magnitude, but the amplitude is twice weaker than the
observed Phillips relation.  Thus in the present model, the
orientation effects can be considered as a possible source of
dispersion in the underlying relation, at the level 0.12~mag.
\item The intrinsic mild asymmetry of the model density structure, 
leads to a continuum polarization at maximum light of $P \la 0.1 \%$,
somewhat lower than that typically observed.  This may indicate that
other effects not included in the present model (e.g., rotation) may
be also required to reproduce the large scale ejecta asymmetry.
\item The model shows line polarization up to a level $P\sim 2\%$ 
in the Si~II ``6150'' and Ca~II IR-triplet features, due to the
aspherical distribution of IMEs.  The line polarization is large from
inclinations where the line velocities are highest, resembling the
observed behavior noted in \cite{Leonard_SNIa}.  However significant
line polarization ($P \sim 1-2\%$) can also be seen from some viewing
angles where the line velocities are relatively low, a characteristic
that as of yet has no correspondence in observations.
\item In the model, the material synthesized during the initial
deflagration phase is transformed into a high velocity ($v \sim
25,000-35,000~\kms$) IME-rich outer shell.  This shell resembles the
structure considered in \cite{Kasen_GCD}, but in the present case has
a larger covering factor and a more irregular structure.  Although not
considered directly in this paper, this material offers one
possibility to explain the detached, high-velocity absorption
features commonly observed in early time SN~Ia spectra
\citep{Mazzali_HVF}.
\end{itemize}

The comprehensive methodology applied in this paper helped to
appreciate certain deficiencies in the theoretical calculations.  The
failure to reproduce the Na~I features in the post-maximum spectra
follows from the use of an approximate alpha-network nucleosynthesis
scheme lacking sodium.  Similarly, the lack of calcium in the outer
high-velocity IME shell is a direct consequence of the crude nuclear
burning scheme used during the deflagration phase.  The wrong trends
noted at later time ($\texp > 60$~days) model colors likely indicate a
breakdown of the assumptions adopted in the radiative transfer
calculations, in particular the neglect of non-thermal ionization
effects.  Meanwhile, the poor correspondence in the model $H$- and
$K$-band light curves with observations reflects, in large part, the
inadequacy of the available atomic line list data.

The current Y12 model is only one example of the family of DFD
explosions (Paper~I).  Other explosion models from that family show
variations in the total \Nifs\ mass, explosion energy, morphology and
ejecta dynamics.  With the addition of even more advanced physics, it
is natural to expect that interesting new behaviors will emerge.
Using a more realistic progenitor structure and ignition conditions
may affect the nucleosynthesis in the model, thereby altering the
composition of the core region, energetics of the deflagration, and
overall explosion energy.  That in turn is likely to influence certain
model observables, such as the absolute luminosity, the dispersion in
$B$-band decline rates, and the model continuum polarization.

A major advance of the multi-dimensional radiative transfer
calculations presented in this paper is our ability to quantify the
dependence of model observables on the viewing angle.  We identified
several intriguing orientation effects in the Y12 model which suggest
specific ways in which the asphericity of SNe~Ia may contribute to
their photometric and spectroscopic diversity.  In particular,
asymmetry in the velocity distribution of iron group elements leads to
a strong dependence of the $B$-band decline rate on viewing angle.
This fact both constrains the degree of compositional asymmetry
present in the ejecta of SNe~Ia and points to a potential source of
dispersion in the SN~Ia width-luminosity relation.  Although not the
case for the Y12 model, one may be tempted to wonder whether a
multi-dimensional SN~Ia model could be imagined such that the peak
luminosity variations of SNe~Ia and their correlation with the light
curve width might arise, at least to some degree, from an orientation
effect.

The comparison of SN theory and observation faces new and interesting
challenges once multi-dimensional models are considered.  Given the
viewing angle dependence, the predictions of aspherical models bear an
intrinsic multiplicity.  Model validation can then no longer be
limited to the traditional exercise of matching synthetic light curves
and spectra to individual SN observations.  Rather, we must also study
the probability distributions and dispersion levels characterizing
various model observables (e.g., peak magnitudes, decline rates, line
velocities and polarization levels) along with the internal
correlations relating different sets of such observables.  These
statistical properties of the model can then be compared to those of a
relevant sample of observed SNe.  Although we have touched on this
methodology in several places here, more generally the situation will
be further complicated by the presence of additional dimensions of
diversity other than asymmetry alone.  For example, the DFD and other
theoretical paradigms admit a family of SN~Ia models which range in
\Nifs\ mass.  More extensive validation
studies thus await further specificity of the theory.  Equally
important will be the development of massive and easily accessible
observational databases containing light curves, spectra, and
polarization for a large number of SNe.

The general success of the Y12 model in reproducing the basic
properties of observed SNe~Ia is not entirely surprising, considering
that the DFD scenario represents a variation on the standard 1D
delayed-detonation models which have been shown to be in fair
agreement with observations
\citep[e.g.,][]{Hoeflich_Khokhlov_LC, Hoeflich_94D}.  
Moreover, the ejecta asymmetry studied here may be generic to a
broader class of SN~Ia explosion models. Standard delayed-detonation
models, for example, show a similar systematic offset in the ejecta
compositional structure if the transition to detonation is triggered
at one point off-center
\citep[]{Livne_OffDD, Gamezo_DD}.  In such models, nuclear
burning extends to higher velocity on the side of the ejecta where the
detonation takes place, just the opposite of that seen in the Y12
model. Nevertheless, the overall morphologies are similar, and the
models likely share many of the same photometric/spectroscopic
variations and polarization signatures discussed here.

Ultimately, the viability of the DFD explosion scenario is subject to
further tests of the robustness of the DFD mechanism itself once the
simplifying assumption of axial symmetry is removed and when more
realistic initial ignition conditions are considered.  Nevertheless,
the overall good agreement of the Y12 model with a broad range of
SN~Ia observations certainly warrants future, more in-depth
investigations into the DFD and related multi-dimensional
delayed-detonation scenarios.  This agreement is especially impressive
given that the model was not artificially manipulated in order to
match observations, rather the system was evolved without intervention
from the initial ignition to several months past maximum light.  We
believe that this sort of end-to-end methodology is critical in
developing and validating future theoretical models of thermonuclear
supernova explosions.

\acknowledgements
DK is supported by the Allan C. Davis fellowship at Johns Hopkins
University and the Space Telescope Science Institute. This work is
supported in part by the U.S. Department of Energy under Grant No.\
B523820 to the Center for Astrophysical Thermonuclear Flashes at the
University of Chicago.  This research used resources of the National
Energy Research Scientific Computing Center, which is supported by the
Office of Science of the U.S. Department of Energy under Contract
No. DE-AC03-76SF00098.


\end{document}